\documentclass[12pt]{article}

\usepackage{caption, subcaption, amsmath,amsfonts,amssymb,amsthm, graphicx, authblk}
\usepackage[margin=1in]{geometry}
\usepackage[colorlinks]{hyperref}
\usepackage[numbers,sort&compress]{natbib}

\graphicspath{ {images/} }

\newcommand{\ket}[1]{|{#1}\rangle}

\begin{document}


\title{Preparing Rernomalization Group Fixed Points on NISQ Hardware}

\author[1]{Troy J. Sewell \thanks{tjsewell@umd.edu}}
\author[2]{Stephen P. Jordan}
\affil[1]{Joint Center for Quantum Information and Computer Science, University of Maryland, College Park, MD 20742, USA}
\affil[2]{Quantum Systems, Microsoft, Redmond, WA 98052, USA}

\renewcommand\Authands{ and }

\date{\today}
\maketitle

\begin{abstract}

 Noisy intermediate-scale quantum (NISQ) hardware \cite{Preskill_2018} is typically limited to low-depth quantum circuits to limit the number of opportunities for introduction of error by unreliable quantum gates. A less-explored alternative approach is to repeatedly apply a quantum channel with a desired quantum state as a stable fixed point. Increased circuit depth can in this case be beneficial rather than harmful due to dissipative self-correction. The quantum channels constructed from MERA circuits can be interpreted in terms of the renormalization group(RG), and their fixed points are RG fixed points, i.e. scale-invariant systems such as conformal field theories. Here, building upon the theoretical proposal of Kim and Swingle \cite{DMERA}, we numerically and experimentally study the robust preparation of the ground state of the      critical Ising model using circuits adapted from the work of Evenbly and White\cite{ERwav}. The experimental implementation exhibits self-correction through renormalization seen in the convergence and stability of local observables, and makes essential use of the ability to measure and reset individual qubits afforded by the ``quantum CCD'' architecture of the Honeywell ion-trap\cite{Pino_2021}. We also numerically test error mitigation by zero-noise extrapolation schemes specially adapted for renormalization circuits, which are able to outperform typical extrapolation schemes using lower gate overhead.
\end{abstract}

\section{Introduction}
\label{sec:intro}

The current period of experimentation on digital quantum devices that are still quite limited in terms of size and error rates has been dubbed the Noisy Intermediate-Scale Quantum (NISQ) era \cite{Preskill_2018}, and it is an ongoing question within the field of quantum computation to find what techniques, if any, may be leveraged to implement practical quantum algorithms on these early devices.  Prior to the development of fully fault-tolerant quantum computers, error mitigation procedures will be needed to achieve robust quantum computation in the presence of noise. In \cite{DMERA}, Kim and Swingle proposed that the stable nature of renormalization group fixed points could provide intrinsic robustness to a class of quantum state-preparation algorithms based on a variant of the Multiscale Entanglement Renormalization Ansatz (MERA) that they named Deep MERA, or DMERA. Similar circuits were also considered for their noise-resilience in \cite{Borregard}. MERA are a well-known variational tensor network ansatz for describing ground states of scale-invariant quantum systems, such as at critical points. The expectation values of local observables can be computed in polynomial time classically using MERA. Unfortunately, the polynomial scaling with bond dimension $\chi$ is of very high degree (e.g. $O(\chi^8)$ for 1D and $O(\chi^{16})$ for 2D using the schemes proposed in \cite{EV09}), making MERA impractical for many classical computations. However, a MERA tensor network may be interpreted directly as a quantum circuit for preparing the state described. Thus MERA circuits can be used to calculate local observables on a quantum computer with a polynomial but large scaling advantage relative to classical computers. DMERA is a variant of MERA more directly tailored to quantum state preparation applications by dealing with local quantum circuits of variable depth rather than general unitary tensors of variable bond dimension.

Given a quantum circuit for preparing a quantum state and a subset of the qubits of the state to be prepared, one can define the causal cone of the subset by discarding all qubits and gates at the earliest stage possible such that the reduced density matrix on the chosen subset remains unchanged. A noteworthy property of DMERA circuits is that geometrically local patches selected from one-dimensional DMERA states have causal cones of constant width, independently of the volume of the one dimensional state from which they are taken. Viewed operationally, the causal cone can be regarded as a quantum channel for preparing the reduced density matrix for the local patch of the overarching state. This channel is composed by iterating a procedure in which new qubits are introduced in the $\ket{0}$ state, unitary gates are applied, and some qubits are discarded. From a renormalization group point of view, this procedure can be regarded as zooming in on a smaller region of the underlying scale-invariant system by fine-graining the lattice discretization, and then discarding the qubits that move out of the field of view under this zooming process \cite{DMERA}. The ability to measure local observables and correlation functions by preparing the causal cone of a small subset of qubits from a large state can dramatically reduce the resources needed and effects of noise \cite{shehab2019noise}, both of which are crucial for developing realistic applications for NISQ devices.

The quantum channel and renormalization group flow points of view are related according to $\lambda = 2^{-\Delta}$, where $\lambda$ is an eigenvalue of the superoperator defining the local quantum channel being applied at each iteration that scales the state by a factor of two, while $\Delta$ is the scaling dimension of the associated operator in the underlying conformal field theory\cite{Giovannetti_2008}. If the scaling dimensions are all positive, as is generically the case, then the quantum channel has a unique attractive fixed point toward which any initial state will converge exponentially at a rate bounded by the smallest scaling dimension. This provides the state-preparation procedure with a degree of intrinsic self-correction. Dissipative quantum state preparation and computation has a long history \cite{Verstraete, Beige, BBT00, Zanardi, Kliesch, Kraus, Marshall, Sinayskiy1, Sinayskiy2} and within this context the DMERA perspective adds a systematic way of deriving concrete quantum circuit implementations of the needed quantum channels for a class of state preparation problems as well as an immediate way to predict convergence rates based on known scaling dimensions of conformal field theories. 

Here, we experimentally and numerically investigate the application of DMERA for the preparation of the ground state of the critical Ising model. Via the Jordan-Wigner transformation, this model can be solved using techniques for free-fermion models. Exploiting the structure provided by this exact solution, Evenbly and White have derived wavelet-based analytic expressions for the tensors by which a local free-fermion MERA can describe the ground state \cite{ERwav}. Due to the exactly solvable nature of this model, the circuit ansatz is an example of a matchgate circuit which can be simulated efficiently classically \cite{Jozsa_2008}. We thus do not propose this model for a demonstration of quantum speedup. Rather, by choosing a model with analytically determined parameters in the tensors, we can make a clean investigation of convergence and error-robustness without confounding issues coming from optimization of variational parameters. A modified version of the circuit with additional gates dual to non-quadratic fermion operators would break the integrability of the circuit dynamics. This would lead to structurally analogous circuits that are challenging to simulate classically, and likewise able to prepare non-Gaussian fermionic states which arise in interacting fermion models and nonintegrable spin chains. 

A noteworthy feature of this state preparation method is that it requires the repeated introduction of fresh ancilla qubits in the $\ket{0}$ state and subsequent discarding of qubits that are no longer needed. It is due to this open-system nature of the procedure that entropy can be removed and self-correction can be attained. This is a necessary feature for all dissipative quantum algorithms, as well as for the ``holographic'' methods of \cite{kim2017holographic, FossFeig, Anikeeva}, but is not possible on most present-day quantum computing test beds. However, measurement and resetting of individual qubits during the execution of a quantum circuit is possible on the ``quantum 2'' architecture used by the Honeywell $\textrm{H\O}$ device on which our experiments were performed\cite{Pino_2021}. In principle, the quantum channel derived from DMERA can be repeated indefinitely on a quantum CCD device, such that the attractive nature of the fixed point constitutes an active form of error correction which perpetually steers the quantum state back toward its target. This is in stark contrast with most NISQ algorithms which rely on limited circuit depth to limit the accumulation of error.

\section{Preliminaries}

The DMERA circuits we consider consist of a depth $D$ layer with $D$ variational parameters which is iterated to drive the system into an approximate ground state supported on a limited number of qubits. The circuit appears to be local within one depth $D$ scale transformation. However, if we fix the geometry of the qubits in the final state of the circuit, earlier layers require interaction of pairs of geometrically distant qubits. This fact favors implementation on near-term ion-trap computers where nonlocal gates can be implemented natively without swap gates. Larger-depth implementations of the quantum channel achieve smaller systematic error arising from the limitations of the ansatz, but incur greater amount of error from imperfect implementation because of the increased number of gates in the past causal cone. We explore numerically the tension between approximation accuracy and noise susceptibility to find the optimal channel depth for a given error rate.
This suggests that while the overall depth can be quite large with fixed points stabilized by local dissipative dynamics, NISQ devices will still need to focus on channels with relatively short depth to limit the effect of noise. We find numerically that the approximation accuracy is exponential in $D$ while the noise susceptibility scales linearly with $D$, suggesting an optimal depth such that $\epsilon \sim e^{-D} / D$ for gate error rate $\epsilon$. For large enough $D$ this is just $D \sim \log(1/\epsilon)$, however, for the larger error rates of NISQ devices we will be working in the regime of small depth where subleading terms and constant factors are crucial for choosing the optimal circuit for particular task at hand.

\begin{figure}[htbp]
\centering
\begin{subfigure}[b]{.45\textwidth}
 \includegraphics[width=\linewidth]{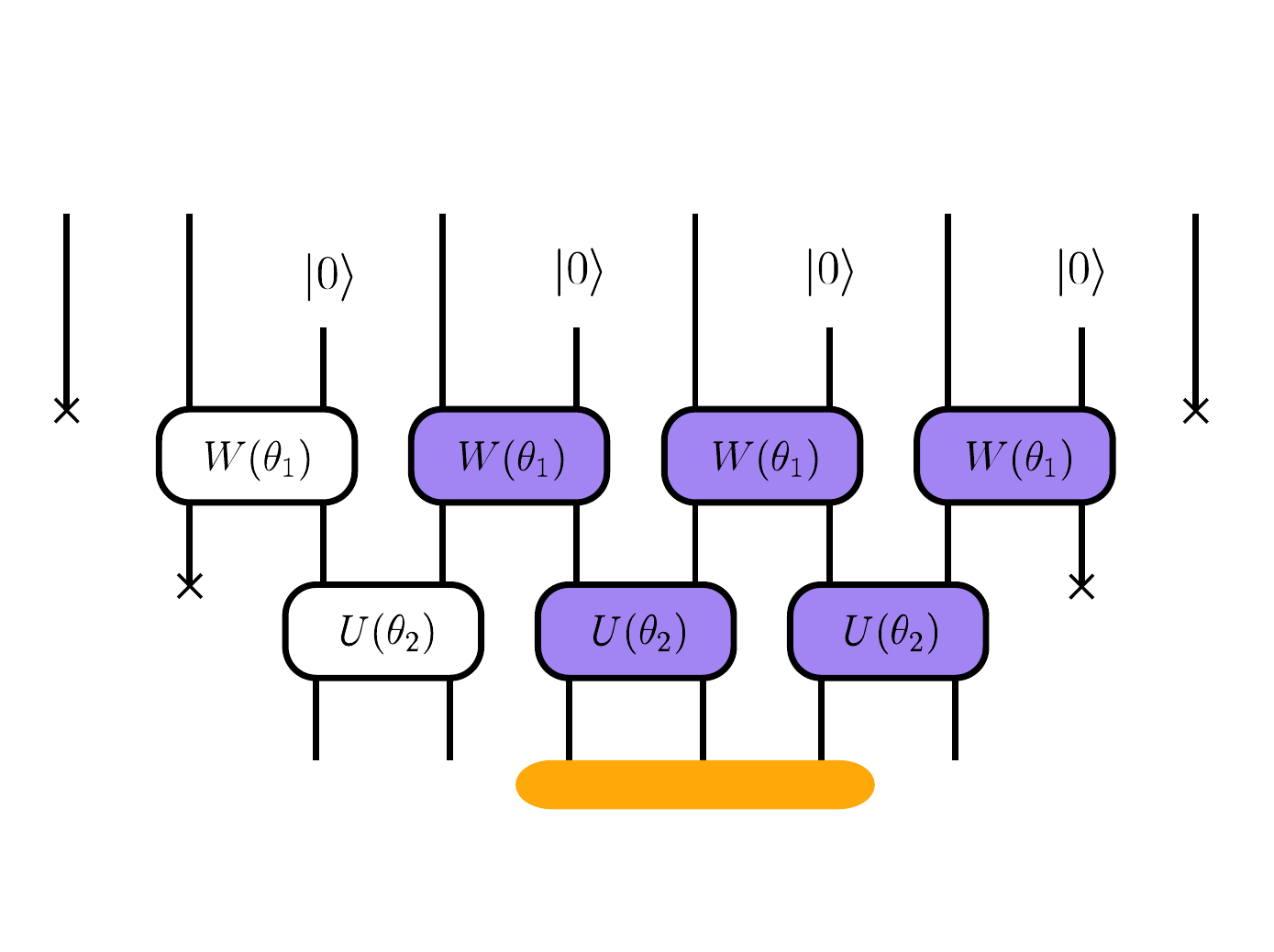}
 \caption{$D=2$ layer with 3 qubit causal cone.}
 \label{fig:D2}
\end{subfigure}
\begin{subfigure}[b]{.45\textwidth}
 \includegraphics[width=\linewidth]{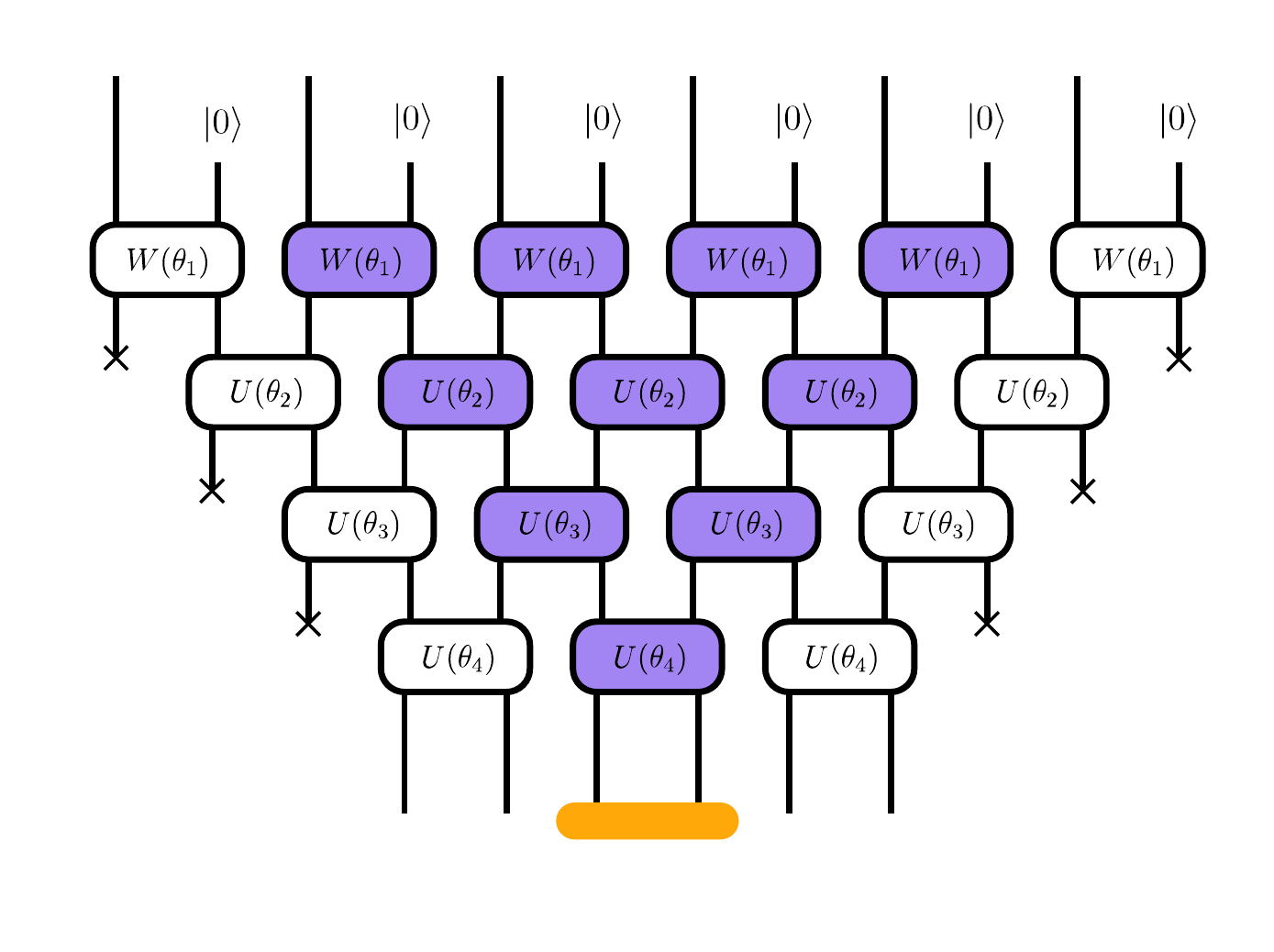}
 \caption{$D=4$ layer with 2 qubit causal cone.}
 \label{fig:D4}
\end{subfigure}
\caption{quantum channels on six qubits performing one scale of entanglement renormalization, with a causal cone highlighted in purple for a local operator }
\label{fig:circs}
\end{figure}

Each layer of the circuit implements one scale transformation by a spatially homogeneous depth $D$ brickwork circuit of two-qubit gates. Within a single scale the local gates are spatially homogeneous, built by one initial layer of "isometry" gates $W(\theta_1)$ and $D-1$ layers of gates $U(\theta_i)$ for depth $i$ within the layer, as seen in Fig. \ref{fig:circs}.

The two-qubit gates $W$ and $U$ are matchgates and therefore preserve parity. They are each parameterized in terms of a single angle, the rotation angle of the even-parity subspace, which gives us $D$ variational parameters for the angle at each depth within the scaling transformation. The odd-parity subspace is fixed by permutation symmetry of the qubits involved in each gate. The unitary gate $U$ is symmetric under the simultaneous permutation of incoming and outgoing qubits, which fixes the action on the odd-parity subspace to be the identity, up to a phase. The $W$ gates are in spirit isometries with only one incoming qubit, since one of the two qubits is always introduced as a $\ket{0}$ state, so they are symmetric under permutations of only the two outgoing qubits. This has the effect of propagating an incoming $\ket{1}$ to the symmetric pair $\ket{01} + \ket{10}$, while an incoming $\ket{0}$ state becomes an arbitrary even-parity state which is always permutation-symmetric. 

\begin{equation}
W(\theta) =\begin{pmatrix}
\cos(\theta-\frac{\pi}{4}) &0 & 0 & \sin(\theta-\frac{\pi}{4})\\ 
0 & \frac{1}{\sqrt{2}} &\frac{-1}{\sqrt{2}} &0 \\ 
 0 & \frac{1}{\sqrt{2}} & \frac{1}{\sqrt{2}} &0 \\ 
 -\sin(\theta-\frac{\pi}{4})& 0 & 0 & \cos(\theta-\frac{\pi}{4})
\end{pmatrix} \hspace{5mm}
U(\theta) =\begin{pmatrix}
\cos(\theta) &0 & 0 & \sin(\theta)\\ 
0 & 1 &0 &0 \\ 
 0 & 0 & 1 &0 \\ 
 -\sin(\theta)& 0 & 0 & \cos(\theta)
\end{pmatrix}
\end{equation}

Since our state is scale-invariant we apply the same scale transformation repeatedly, implementing identical steps of fine-graining to prepare the renormalization UV fixed point defined by our local scaling operation. Since the past causal cone of a local region is constant width, we can use this to define a quantum channel on a constant number of qubits that implements the local scaling operation on a fixed subsystem size. Since each scaling layer introduces new qubits in order to fine-grain the state by zooming in on the virtual lattice, the constant qubit channel also needs to discard qubits at the edges which end up outside of the causal cone. This combination of preparing new ancilla and discarding edge qubits is built into the definition of the local scaling channel along with the unitary scale transformation, making the channel non-unitary but physically implementable on a quantum computer.

Since the unitary circuit implementing the channel is only invariant under translation by an even number of sites, there are two different ways the subregion of retained qubits can align with the circuit. The two choices of subregion are akin to choosing to follow a subregion moving to the left or right when descending down the state preparation circuit of a large spin chain. Sampling over infinite sequences of these left and right channels would correspond to an averaging of the location of our $n$ qubit subregion over the the infiite sized spin chain. 
We implicitly perform this averaging of subregion location by studying the channel which is the statistical mixture of the left and right channels with equal probability, which ensures that the fixed point energies measured are still variational for the infinite sized system. 

\begin{figure}[htbp]
\centering
 \includegraphics[width=.55\linewidth]{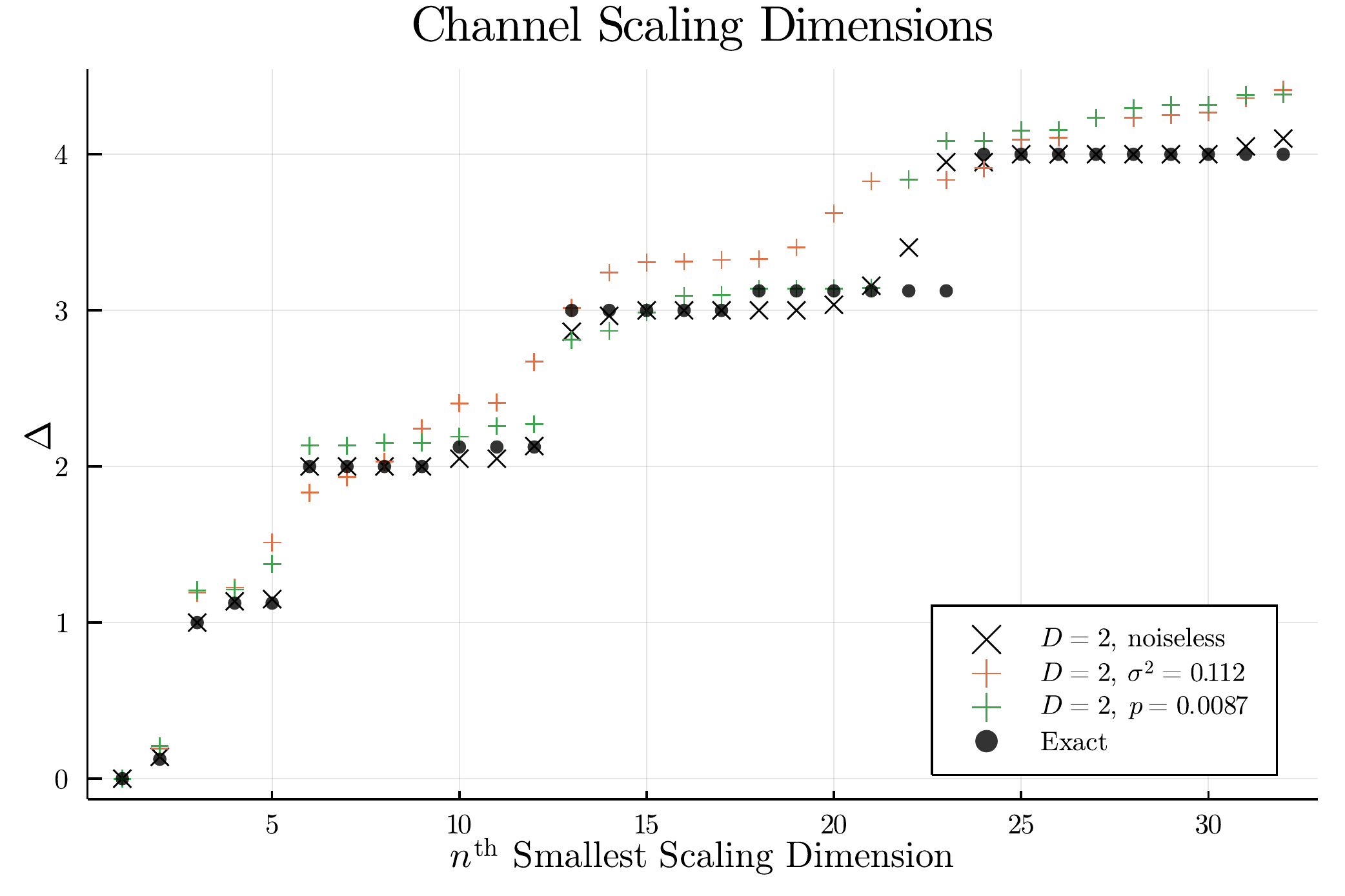}
\caption{Scaling dimensions $\Delta$ extracted from the eigenvalues $\lambda$ of the quantum channel for $D=2$. Exact values for the 32 smallest scaling dimensions of the Ising CFT compared with those derived from the largest 32 eigenvalues of the noiseless channel. Perturbations to these scaling dimensions in noisy versions of the channel are also included.}
\label{fig:scalingDims}
\end{figure}

There is a single eigenvalue of the quantum channel with $\lambda = 1$, corresponding to the fixed point operator of the channel. All other operators have eigenvalues with magnitude less than one, meaning those operators will decay exponentially with successive iterations of the channel. The scaling dimensions of these operators are given by $\Delta = - \log_2 \lambda$. 
A local $n$-qubit channel has $2^{2n}$ eigenvalues, the largest of which correspond to the slowly decaying low energy modes described by the CFT with small scaling dimension. 
We compare the scaling dimensions of the primary operators of the CFT describing the critical Ising model with those derived from the local quantum channels, seen in Table \ref{tab: sca-dims}. Additional scaling dimensions are shown in Fig. \ref{fig:scalingDims}, where the smallest 32 scaling dimensions in the CFT are compared with those derived from the largest 32 eigenvalues of the $D=2$ channel on three qubits.
The largest eigenvalues match well with the expected scaling dimensions, suggesting that the low energy sector of the spin-chain is indeed well described as a CFT, while the smallest eigenvalues that correspond to rapidly decaying high energy modes that are not universal and do not correspond to quantities of the CFT. 

The positive scaling dimensions and unique fixed point imply that local subsystems converge exponentially and only $\log(1/\epsilon)$ layers are needed to prepare the target state with precision $\epsilon$. However, even the exact fixed point is still just an approximation of the true ground state, limited in precision by the depth $D$ of the ansatz used, and the actual fixed point state prepared by a NISQ device will have perturbations from the ideal fixed point introduced by noise.

\begin{table}[h]
\begin{center}
\begin{tabular}{c|c|c|c}
 & exact & $D=2$& $D=4$ \\ \hline
 $E^*/L$ & -1.27323.. & -1.23948.. & -1.26757.. \\\hline
 $\Delta_I $ & 0 & 0 & 0 \\\hline
$\Delta_\sigma $ & 0.125 & 0.136.. & 0.123.. \\\hline
$\Delta_\epsilon $ & 1 & 1 &1 \\\hline
\end{tabular}
\caption{Critical data extracted from quantum channels preparing approximations of the critical Ising ground state, using wavelet-derived angles of \cite{ERwav} compared with exact values. The exact ground state energy density in the thermodynamic limit $E_0/L = -4/\pi$ is compared to the fixed point state. The scaling dimensions of the three primary operators listed above are derived from the three largest eigenvalues of the quantum channel. Additional dimensions included in \ref{fig:scalingDims}. }
\label{tab: sca-dims}
\end{center}
\end{table}

We can understand why the DMERA circuit based on wavelets is able to prepare the critical Ising state by considering the action of the circuit in terms of fermion modes via a Jordan-Wigner transformation. The critical Ising model spin Hamiltonian $H_I = -\sum_i X_i X_{i+1} + Z_i$ is transformed into the majorana Hamiltonian $H_M = -i \sum_i \gamma_i \gamma_{i+1}$ \footnote{For finite size, the 1D Ising model with periodic boundary conditions is equivalent to a free-fermion model with anti-periodic boundary conditions in the even parity subspace and periodic boundary conditions in the odd parity subspace. We focus on local subregions and observables of an infinite system where these subtleties may be ignored.}, both of which have ground state energy density $E_0/L = -4 / \pi$. Being translation invariant, the eigenmodes of the massless free-fermion Hamiltonian dual to the critical Ising model are plane-waves. Matchgate circuits implement linear transformations on fermion modes, which would require depth scaling linearly with the system size to implement an exact transformation mapping local modes to the proper plane-wave modes via a local quantum circuit. This can be cast as a QAOA circuit with depth $L/2$ with $O(L^2)$ gates to prepare the exact critical Ising ground state for a size $L$ system \cite{Ho_2019}. The wavelet transformation offers an alternate basis of local wave-packets, which can be mapped to from local modes using a $O(D\log ( L))$ depth circuit with $O(DL)$ gates. Since the ground state is invariant under transformations that do not mix the occupied and unoccupied modes, it is conceivable that these wave-packet modes could be made from only occupied modes and therefore prepare the exact ground state. In practice the quasi-local modes found through variational optimization of a low-depth DMERA circuit architecture only prepare an approximate ground state, albeit one whose accuracy increases exponentially with the layer depth $D$. The analysis of \cite{ERwav} considered the tight-binding model of 1D lattice of spinless fermions at half-filling, which can be decoupled into two copies of the Hamiltonian $H_M$ with local transformations. They designed the circuit and set of angles to specify a wavelet transformation on the fermion modes that occupies the Dirac sea of momentum modes with $|k| < \pi /2$, with vanishing support at the Fermi points where $|k| = \pi/2$. Larger-depth variations of the circuit then correspond to higher-order wavelets which are able to prepare a better approximation by canceling higher derivatives of the mode occupation around the Fermi points. These circuits can be directly adapted for the critical Ising model as outlined in the Appendix of \cite{ERwav}, which serves as the basis of our analysis. Additional work on wavelet circuits has been done exploring application to both free-fermion and free-boson systems \cite{PhysRevX.8.011003, witteveen2019quantum, Witteveen_2021}.

The relation of different wavelet basis states by both translations and scaling transformations makes them especially suited for critical models, which have these operations as symmetries. Both the wavelet circuits and MERA tensor networks reflect these symmetries in the spatial homogeneity of tensors at a given depth in the circuit, and the homogeneity of the different circuit layers in the hierarchical structure. This discrete scaling symmetry allows us to interpret the past causal cone of a local region as the iteration of a single quantum channel, and the translation invariance helps to reduce the number of variational parameters needed to find good circuits.

Due to the inherently discrete nature of qubits, the continuum of a field theory and its symmetry group must be discretized and broken to be represented on digital quantum hardware. Additionally the staggered structure of the individual quantum gates used to prepare the state and the fixed scaling by two for each circuit layer break translation and scale invariance of the unitary used to prepare a pure quantum state. These result in small deviations from exact translation invariance and power-law scaling of the correlation functions for the resulting ground state. These deviations can be improved by averaging over different samples of the state transformed by these different symmetries. This can be done implicitly by equally sampling over the past causal cone from different locations of the global state, or symmetrizing the measurement data by reflections or the global parity symmetry.

We find that the wavelet-based gates derived in \cite{ERwav} admit particularly efficient decomposition in terms of M{\o}lmer-S{\o}rensen (MS) gates, the native two-qubit gate set on ion trap devices, which we will sometimes denote $XX(\theta)$. This conveniently allows the implementation of these circuits on NISQ ion-trap computers with little gate overhead. MS gates can typically be applied to any pair of qubits in an ion-trap and so the qubit architecture within a trap is thought of as globally connected. Ion-trap computers engineer this all-to-all connectivity in different ways, for instance by moving individual ions around to bring pairs into interaction sites in the Honeywell quantum CCD architecture \cite{Pino_2021}.

\begin{equation}
 X_iX_j(\theta) = e^{-i\frac{\theta}{2} X_i X_j}
\end{equation}

\begin{figure}[htbp]
\centering
 \includegraphics[width=.48\linewidth]{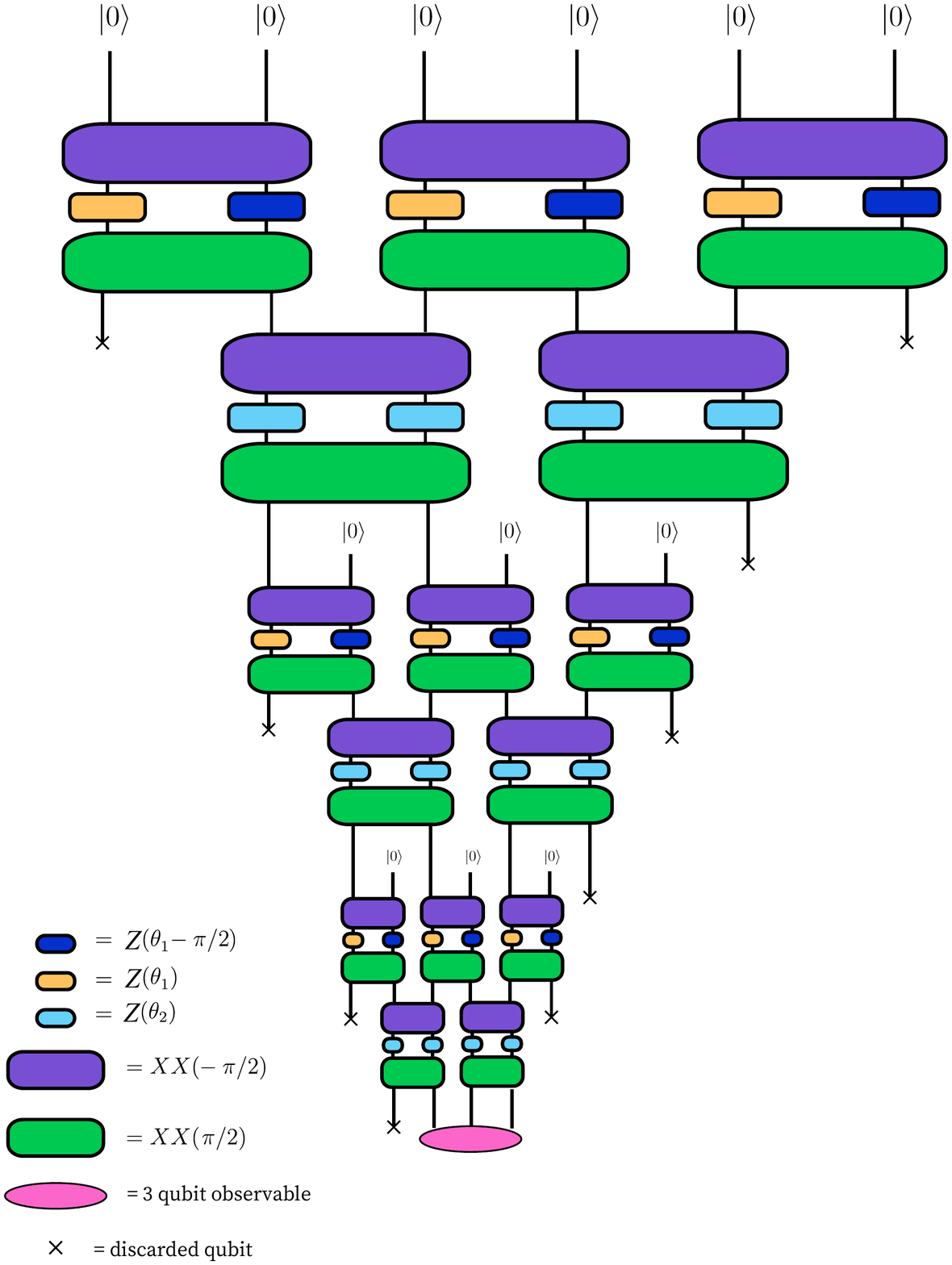}
 \includegraphics[width=.48\linewidth]{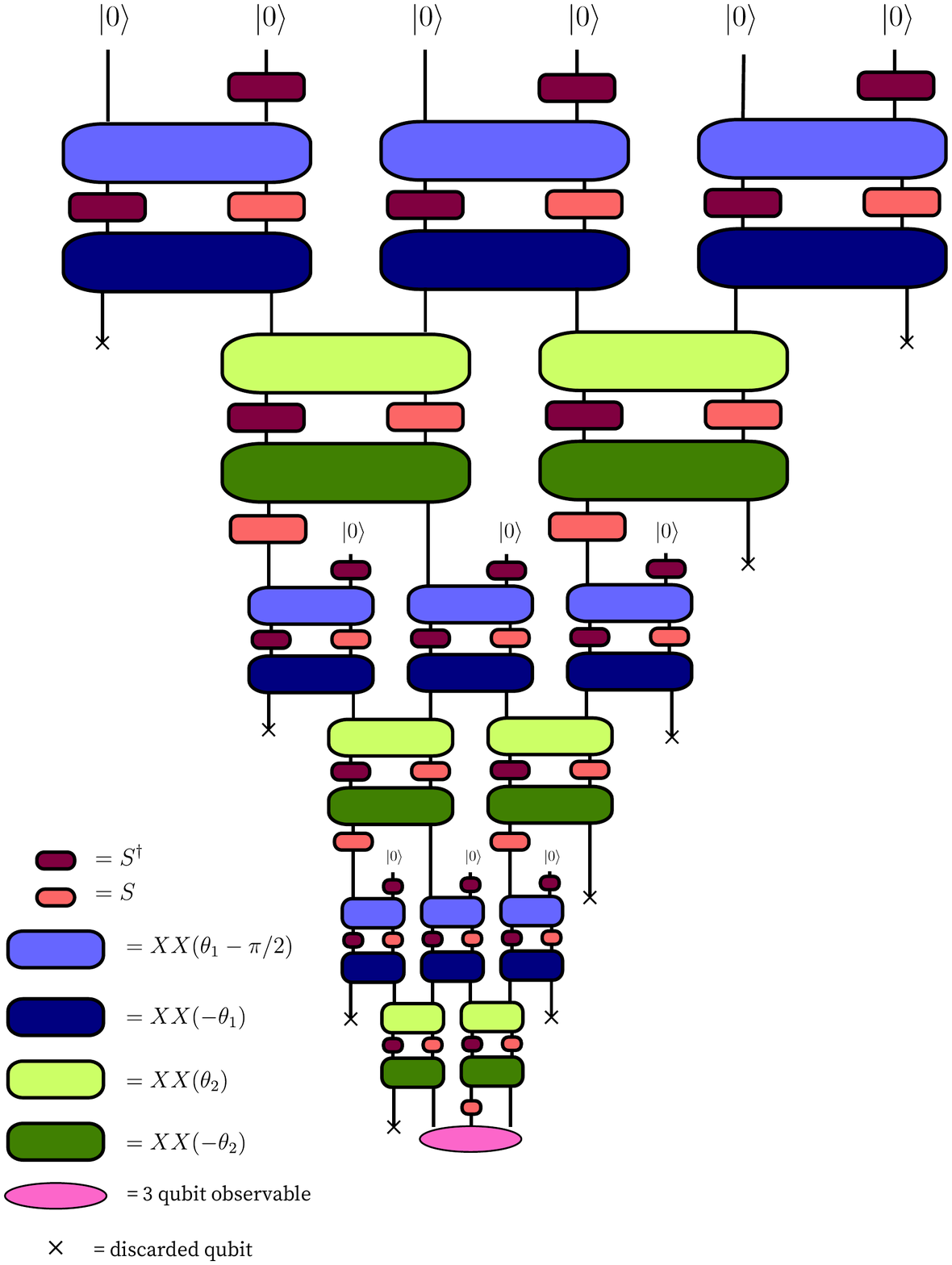}
\caption{Quantum circuit for causal cone of a three qubit observable using three layers of the $D=2$ circuit ansatz starting from the all zeros state, presented in two alternate gate decompositions, $\mathcal{C}_1$ (left) and $\mathcal{C}_2$ (right), both constructed with two-qubit M{\o}lmer-S{\o}rensen gates and single qubit Z rotations.}
\label{fig:3layer}
\end{figure}

The XX native interaction affords a simple implementation of our state preparation circuit, with each gate $W$ and $U$ being decomposable in terms of two MS gates and additional single qubit Z rotations $Z(\phi) = \exp(i \frac{\phi}{2} Z)$. All of our gates are matchgates and thus cannot form a universal gate set in a geometrically local circuit. The addition of local $X$ or $Y$ rotations into the circuit would expand the gate set to become universal. The non-local interaction of these gates is also crucial for implementing a hierarchical algorithm like DMERA, where new qubits need to be interleaved at each layer and would be costly to achieve with swap gates. Importantly, the non-local gates are always performed over strings of qubits in the all zero state, which means they can be performed with local matchgates and fermionic swap gates, so they can still be considered local matchgate circuits that are classically simulable. Local $Z$ rotations are often implemented virtually via the internal clock of the hardware Hamiltonian, making them practically noise free. We considered two possible native gate decompositions of $W$ and $U$, denoted $\mathcal{C}_1$ and $\mathcal{C}_2$, drawn out in Fig. \ref{fig:3layer}.

\begin{align}
 \begin{split}
 W_{\mathcal{C}_1}(\theta) &= XX(\pi/2 )\cdot Z_1(\theta) \cdot Z_2(\theta - \pi/2) \cdot XX(-\pi/2)\\
 U_{\mathcal{C}_1}(\theta) &= XX(\pi/2 )\cdot Z_1(\theta) \cdot Z_2(\theta ) \cdot XX(-\pi/2)\\
 W_{\mathcal{C}_2}(\theta) &= Z_1(\pi/2)\cdot XX(-\theta )\cdot Z_1(-\pi/2) \cdot Z_2(\pi/2) \cdot XX(\theta-\pi/2)\cdot Z_2(-\pi/2)\\
 U_{\mathcal{C}_2}(\theta) &= Z_1(\pi/2)\cdot XX(-\theta )\cdot Z_1(-\pi/2) \cdot Z_2(\pi/2) \cdot XX(\theta)\cdot Z_2(-\pi/2)
 \end{split}
\end{align}

To truly prepare an approximation of the critical Ising ground state, we would need to apply the transitional layer of $XX_{\pi/2}$ gates to every pair of nearest-neighbor qubits after the layers of DMERA circuit have been performed \cite{ERwav}. Instead, we directly target the ground state of the Hamiltonian by applying the transition layer to the usual Ising Hamiltonian $H_I$, resulting in $H_{I^*} = \sum_i -X_i X_{i+1} + X_{i-1}Z_i X_{i+1}$.

\section{Error Models}

The largest single source of error is two-qubit gate infidelity, which we model by imprecise rotation angle of MS gates given by a Gaussian distribution with standard deviation $\sigma$. This noisy version of the MS gate, denoted $\widetilde{XX}$, is equivalent to a quantum channel with Kraus operators $K_0 = \sqrt{\frac{1+e^{- \sigma^2 /2}}{2}} XX(\theta)$ and $K_1 = \sqrt{\frac{1-e^{- \sigma^2 /2}}{2}} XX(\theta-\pi)$, as shown in Eq. \ref{eq: XX}. This is equivalent to a probability of $(1-e^{-\sigma^2 /2})/2$ to flip both bits after the ideal MS gate has been applied, which for small error is $\approx \sigma^2 / 4$ to leading order.

This kind of noise is correlated between qubits, since the bit flips only occur in pairs, and biased towards bit flip errors rather than other two-qubit errors due to its source in gate imprecision.

\begin{align}
\begin{split}
 \widetilde{XX}_{\theta,\sigma}(\rho) &= \int^\infty_{-\infty}d \phi \frac{e^{-(\phi-\theta)^2/2\sigma^2}}{\sqrt{2 \pi \sigma^2}}XX_\phi \rho XX^\dagger_\phi \\
 &= \left(\frac{1+e^{- \sigma^2 /2}}{2} \right) XX_\theta \rho XX^\dagger_\theta + \left(\frac{1-e^{- \sigma^2 /2}}{2}\right) XX_{\theta-\pi} \rho XX^\dagger_{ \theta-\pi}\\
 &= \left(\frac{1+e^{- \sigma^2 /2}}{2} \right) XX_\theta (\rho) + \left(\frac{1-e^{- \sigma^2 /2}}{2}\right) XX XX_{\theta} (\rho) XX
 \end{split}
 \label{eq: XX}
\end{align}

\begin{equation}
\begin{split}
 \mathcal{D}_{p}(\rho) &= (1-p) \rho + \frac{p}{3} \left ( X \rho X + Y \rho Y + Z \rho Z \right)\\
 &= \left(1 - \frac{4p}{3} \right)\rho + \frac{2p}{3} \mathbb{I}_2
 \end{split}
 \label{eq: Dp}
\end{equation}

Spontaneous logical errors from environmental noise can be modeled by periodically applying a single qubit depolarizing channel $\mathcal{D}_{p}$ independently to each qubit, equivalent to applying a Pauli operator randomly with small probability. This noise process is uncorrelated between the qubits and invariant under single qubit rotations. 
The Kraus operators of single qubit depolarizing noise, shown in Eq. \ref{eq: Dp}, are $K_0= \sqrt{1-p}I$, $K_i = \sqrt{p/3}\sigma_i$, for Pauli matrices $\sigma_i$. For the depolarizing error model, a single qubit depolarizing channel with error probability $p/2$ is applied to each qubit before and after the MS gate is applied. 

The depolarizing channel can be made anisotropic by biasing the probabilities for each Pauli operator. This may arise from a particular device coupling to the environment in a way that favors dephasing or bit-flip errors, while still assuming no correlation of the noise process between qubits. Ion-trap devices have relatively long coherence times and so environmental noise of this kind is expected to be smaller than gate implementation error. With these local error models we ignore state preparation and measurement errors, and make many assumptions including that errors are homogeneous in space and time, lack long-range correlations beyond the local gate associated to the error, and are state-independent.

\begin{figure}[htbp]
\centering
 \includegraphics[width=.55\linewidth]{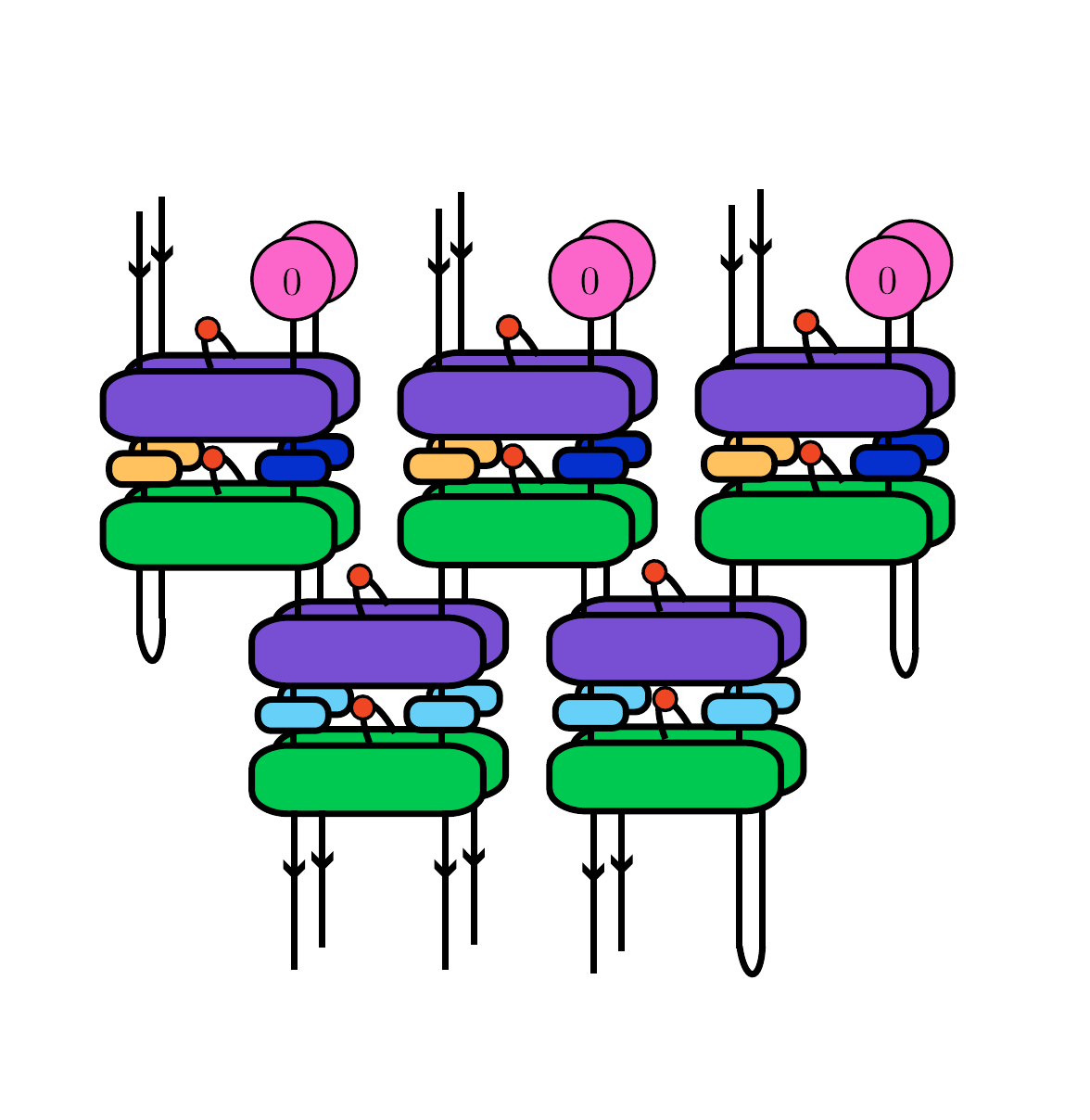}
\caption{Tensor network diagram for the $\mathcal{C}_1$ circuit with $D=2$ on three qubit density matrices. Two copies of the unitary circuit are given to act on each set of indices of the density matrix tensor, with initialized zero states given by one-index tensors, and the discarded qubits given by connecting the dangling legs from each circuit copy. Small red two-index tensors are also given for each MS gate which stores the weights of the Kraus operators which are summed over to implement a given local noise model for the MS gate at that position in the circuit. The colors of all the single and two-qubit gates are chosen to be identical for identical rotation angles. The choice of open indices at the bottom of the diagram make this an instance of the left channel. The left channel is averaged with the right channel, constructed by choosing right three qubits as output, to form the variational channel that implicitly averages over subregion location in the spin chain.}
\label{fig:chanTN}
\end{figure}

The leading term in the average gate infidelity $1-\overline{\mathcal{F}}$, where $\overline{\mathcal{F}}$ is the average gate fidelity, scales as $\sigma^2$ and $p$ respectively for the angle imprecision and depolarizing models. These will be taken as the relevant noise parameters $\epsilon$ for their respective error models. We also expect the fixed point infidelity and local observable expectation values to scale linearly in these parameters to leading order, which is confirmed by our numerics. The coefficient of this scaling is determined by specific details of the noise models and circuit implementation, including native gate sets and choices of gate decomposition.
The tensor network for the channel acting on the density matrix of three qubits is given in Fig. \ref{fig:chanTN} with the $\mathcal{C}_1$ gate decomposition in terms of native $XX$ and $Z$ gates.

\begin{figure}[htbp]
\centering
 \includegraphics[width=.6\linewidth]{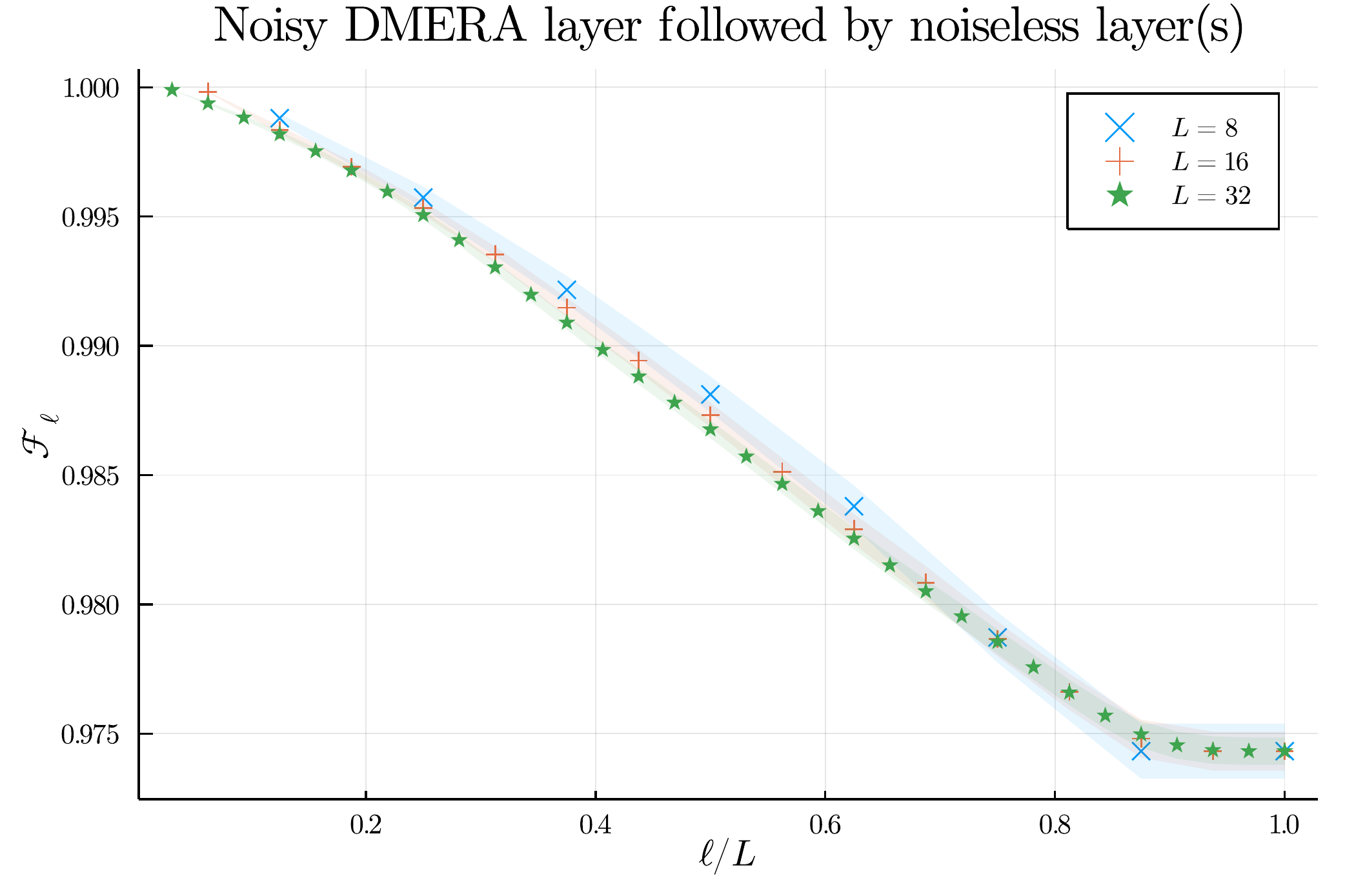}
\caption{Average fidelity $\mathcal{F}_\ell$ of size $\ell$ subsystems between ideal DMERA circuit preparing $L$ spin ground state and states where the initial layers have imprecise angles. As successive layers dilute early circuit errors subsystem fidelities remain roughly constant as a fraction of total system size, i.e. $\ell / L$.}
\label{fig:subFidN}
\end{figure}

\begin{figure}[htbp]
\centering
 \includegraphics[width=.55\linewidth]{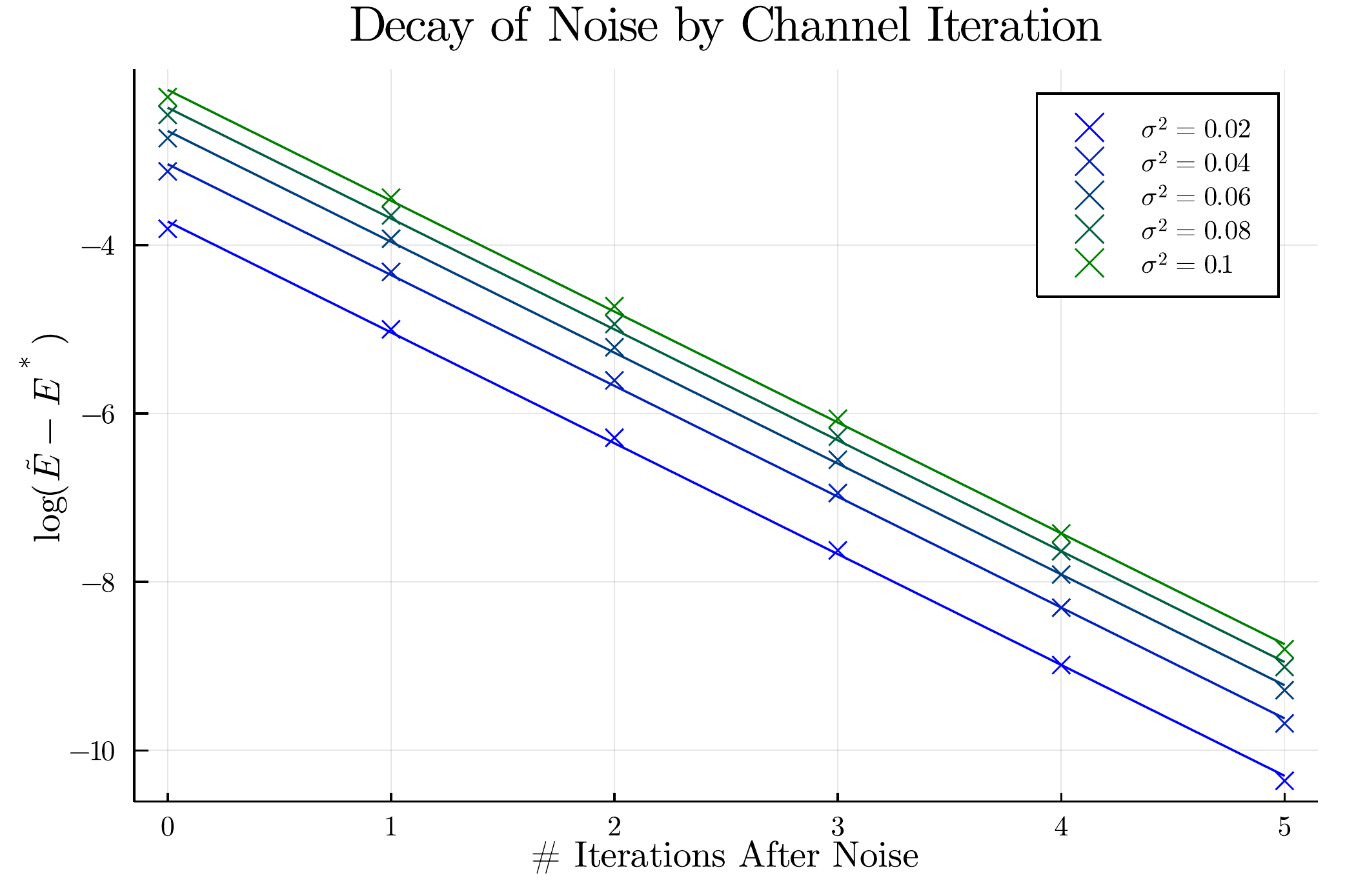}
\caption{Response of the noiseless $D=2$ channel to noisy gates,the channel with $\sigma^2$ noise is applied once to the noiseless fixed point state. The difference between the noisy energy density $\tilde{E}$ after successive applications of the noiseless channel and the fixed point energy density $E^*$ is plotted on a log scale, showing an exponential convergence back to the fixed point energy density. }
\label{fig:noiseResponse}
\end{figure}

We see in Fig. \ref{fig:subFidN} that errors from a prior layer tend to dilute evenly among the qubits in subsequent layers. Here we apply one noisy scale transformation to a global state on four bits, followed by multiple noiseless circuit layers which each double the system size of the ground state on a circle. The global state fidelity with the reference pure state prepared by the noiseless circuit is constant as more layers are applied, and the subsystem entropy stays roughly constant. Since prior errors mix and stabilize across the system as the total size grows, the error on a constant size subsystem decreases exponentially as more layers are applied. An individual gate error introduces a perturbation to the fixed point state, which may be decomposed into the channel eigenoperators. Each operator perturbing the fixed point state decays exponentially according to its eigenvalue, which means the true convergence of an expectation value to the asymptotic value is a sum of exponential decays. In Fig. \ref{fig:noiseResponse} we see that this can be fitted well by a single exponential with a rate of convergence back to the fixed point value is given by a scaling dimension of $\Delta \approx 1.89$. This convergence rate is consistent across different noise strengths in the gate imprecision model.

The stability of this channel against noise can be thought of in terms of an equilibrium reached by the noisy channel by discarding error-laden qubits and introducing new high fidelity qubits each layer. By introducing new qubits, the average number of gates a given qubit has seen remains low. By discarding qubits, a prior error may end up entirely on the discarded qubits, or in the phase of the entanglement between the discarded qubits, both of which would have no effect on observables or fidelities of the subregion we keep. 

Given that an $n$-qubit channel has $O(nD)$ gates that each have a gate error $\epsilon$, and the errors decay with average eigenvalue $\lambda$ under action of the channel, we can expect the total error of an $n$-qubit fixed point to be $\delta \sim nD\epsilon + nD \epsilon \lambda + nD \epsilon \lambda^2 + \ldots = nD\epsilon/(1-\lambda)$.

\begin{figure}[htbp]
\centering
 \includegraphics[width=.55\linewidth]{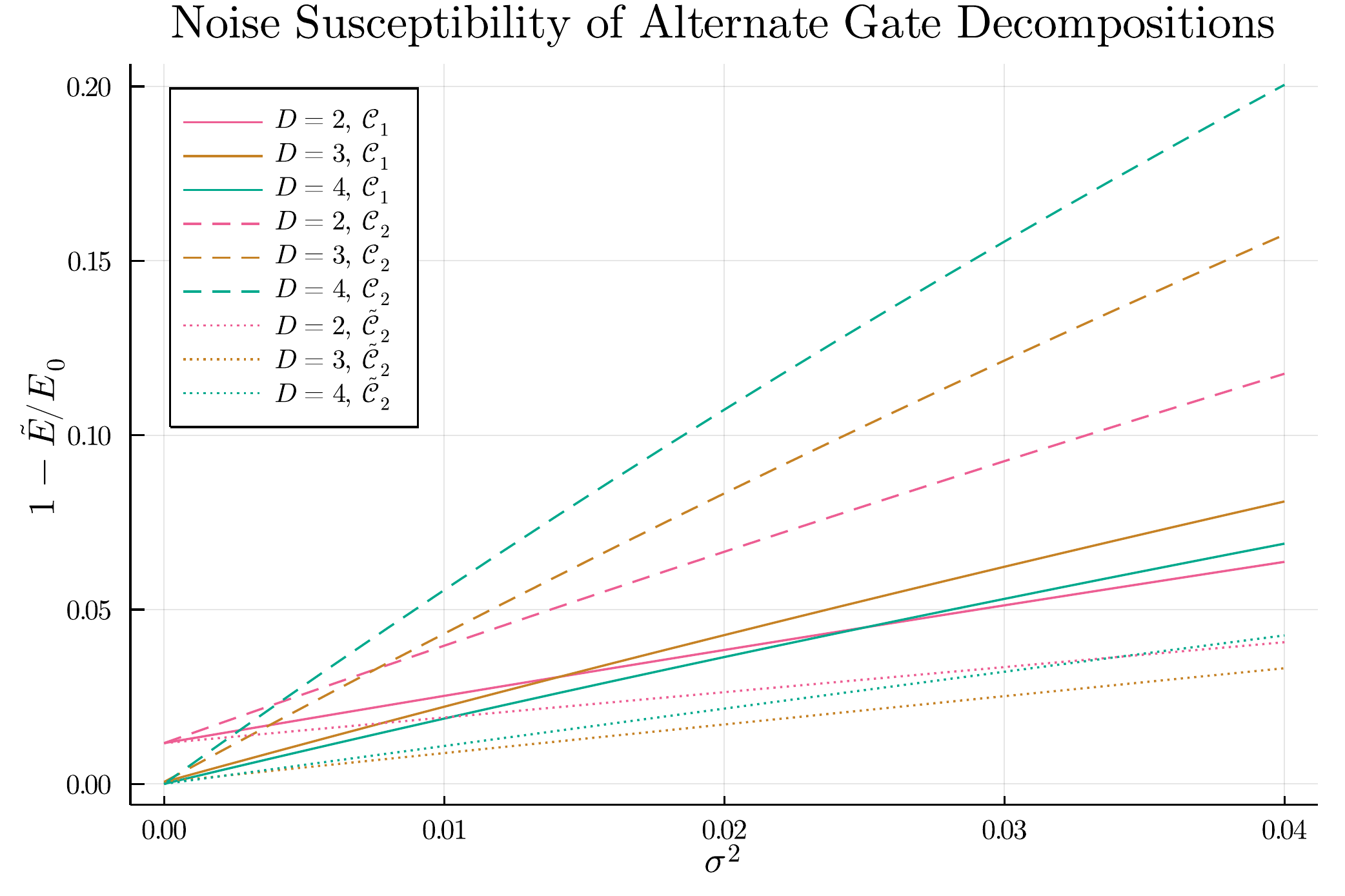}
\caption{Percent error of fixed point energy densities for $D=2,3,4$ channels using the angle imprecision error model in the $\mathcal{C}_1$(solid) and $\mathcal{C}_2$ (dashed) gate decompositions. A modified model where angle variance scales with the target rotation angle was also tried for $\mathcal{C}_2$ (dotted), but would have no effect on $\mathcal{C}_1$ due to fixed MS rotation angles.}
\label{fig:decomp}
\end{figure}

We do find numerical support that the energy density does scale to leading order in $\sigma^2$ and $p$ for our respective error models. In Fig. \ref{fig:decomp}, which shows the scaling with gate imprecision, we also see that the different choices of gate decomposition can greatly affect the scaling of the noise, with $\mathcal{C}_1$ scaling better than $\mathcal{C}_2$. Interestingly, we also see that the $D=4$ channel unexpectedly scales better than the $D=3$ for the $\mathcal{C}_1$ gate decomposition, but not for $\mathcal{C}_2$. These behaviors were not reflected in the local depolarizing noise model, which had very similar scaling for different decompositions and increasing error susceptibility with depth. This could be due to the uncorrelated and local unitary invariance of the depolarizing noise, which may allow it to propagate in a more regular manner within circuits and avoid the irregularities seen in the gate imprecision model.

In \cite{zhu2020generation} an alternate angle imprecision model is suggested for MS gates, where the Gaussian variance is proportional to the angle of rotation. This angular dependence of the variance would alter the error model of $\mathcal{C}_2$ gate decomposition, but not that of $\mathcal{C}_1$ decomposition used on the Honeywell experiment where the MS gate angles are fixed. Shown dotted in Fig. \ref{fig:decomp}, this model has effectively lower noise since the MS gates have a small rotation angle compared to the fixed angle of $\pi/2$ in the $\mathcal{C}_1$ decomposition.

\begin{figure}[htbp]
\centering
 \includegraphics[width=.55\linewidth]{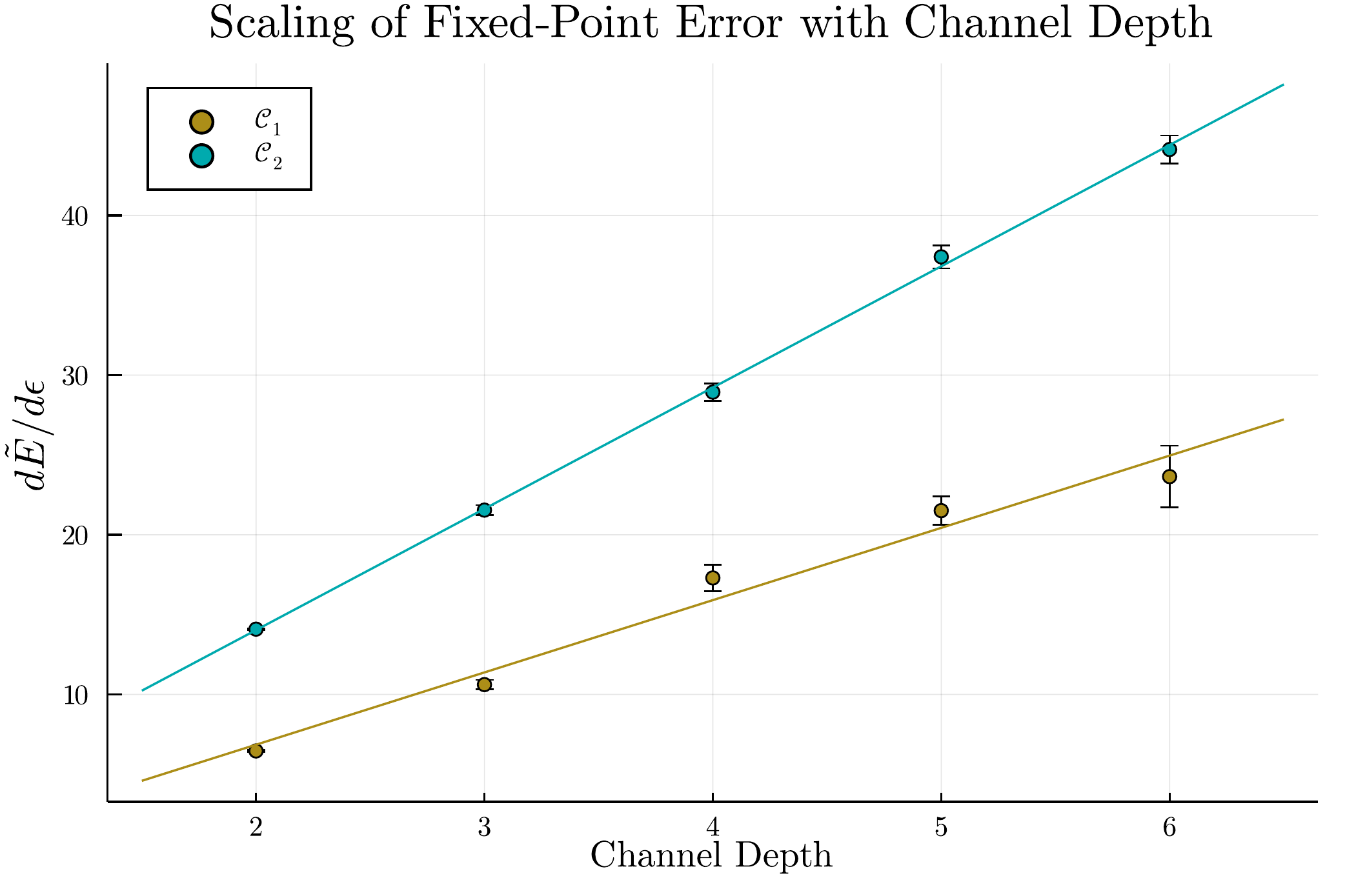}
\caption{Rate of error growth in the fixed point energy density, $d\tilde{E}/d\epsilon$, versus channel depth $D$, using $\mathcal{C}_1$ and $\mathcal{C}_2$ circuit decompositions and the MS gate imprecision model with $\epsilon = \sigma^2$. The error growth rate for each channel depth averaged over many sets of angles with low energy fixed point, suggests a linear growth in error susceptibility with channel depth.}
\label{fig:fixE}
\end{figure}

Since the linear scaling of the energy density in the error parameter tends to increase for larger-depth channels, we compare $d\tilde{E}/d\epsilon$ for channels of different depth to find dependence on $D$ of the fixed point error. To avoid potential peculiarities arising from particular sets of angles having an atypical noise susceptibility, as seen with $\mathcal{C}_1$, we averaged $d\tilde{E}/d\epsilon$ over many sets of angles for fixed depth $D$ in order to assess the typical behavior of channels of a given depth. The different circuits sampled, however, were chosen so that they have a fixed point energy density less than $-1.2$. This constraint is needed because we are only interested in the behavior of circuits that provide reasonable ground state approximations, and $d\tilde{E}/d\epsilon$ can be negative near $\epsilon = 0$ for channels with higher energy fixed points. After averaging the noise susceptibility $d\tilde{E}/d\epsilon$ over different channels of the same depth with low energy fixed point states, we find in Fig. \ref{fig:fixE} that the circuit decomposition $\mathcal{C}_2$ does typically have worse scaling than $\mathcal{C}_1$ under the gate imprecision model, and the fixed point energy of both circuits seem to scale linearly with channel depth $D$.

\section{Error Mitigation}
 
Another kind of error modeling that is important for error mitigation is a functional form to model the effect that error processes have on measurement expectation values. The error mitigation technique of zero-noise extrapolation works by assuming a functional form with certain free parameters, one crucially being the noise-free expectation value, and solving for these parameters by measuring the expectation value at different values of the noise parameters \cite{errorMit}. In practice the noise parameters are not able to be freely tuned, so what is often done is that certain noise sources are amplified by inverting and repeating parts of the circuit $n$ times, $U(U^\dagger U)^n$, so that the source of noise has been applied $2n+1$ times to a logically equivalent circuit. The repeated unitary elements could be any unitary subsection of the circuit, from individual gates to large sections of the circuit. Since there are many methods of experimentally amplifying error processes and functional forms to model the effects of errors on observables, many zero-noise extrapolation schemes are possible with varying success.

In our numerical simulations of zero-noise extrapolation, gate repetitions are implemented on individual MS gates. This could be performed everywhere in the circuit, but would triple the total MS gate count for the lowest order extrapolation. Since the DMERA channel acts to dissipate perturbations from the fixed point state, errors from earlier layers of the circuit are exponentially suppressed as seen in Fig. \ref{fig:noiseResponse}. To reduce the overhead of inserting extra gates into the entire circuit getting the most of error extrapolation we can focus on amplifying and subtracting noise from the final layers of the circuit which have a greater impact on the measured expectation value.

One way to model noise on an observable is as a polynomial perturbation. The measured energy density $\tilde{E}$ is the noiseless asymptotic value $E^*$, plus a polynomial perturbation in the unknown error parameters $\varepsilon_i$ for some independent sources of error, which we assume contribute independently without interaction to the measured energy $\tilde{E}$. To fit a polynomial with $m$ parameters, a collection of at least $m$ appropriate noise amplification measurements are needed in addition to the unamplified measurement to solve for all the coefficients. 

Chunking the error from each channel application into a single source, we expect the error contribution $\varepsilon$ from each application to be identical, but with error from earlier stages of the circuit reduced by the scaling parameter $\lambda$. This suggests a polynomial model of the error with three free parameters:

\begin{equation}
\tilde{E} = E^* + \varepsilon + \varepsilon \lambda + \varepsilon \lambda^2 + \ldots = E^* + \frac{\varepsilon}{1 - \lambda}
\end{equation}

An alternate functional form for the error is a product of exponentials applied to the target energy. Following the same logic, we get an analogous functional form for the noisy energy as an exponential with three free parameters.

\begin{equation}
\tilde{E} = E^* e^{-\varepsilon} e^{ -\varepsilon \lambda} e^{ - \varepsilon \lambda^2}\cdots = E^* e^{ \frac{-\varepsilon}{1 - \lambda}}
\end{equation}

This multiplicative series model is equivalent to the additive series ansatz when expanding the exponential to first order, since $E^*(1 - \frac{\varepsilon}{1- \lambda} + \dots) \approx E^* - \frac{E^* \varepsilon}{1- \lambda} = E^* + \frac{\varepsilon '}{1-\lambda}$, and also has the correct behavior for large $\varepsilon$ where the energy density should approach zero.
 
\begin{figure}[htbp]
\centering
 \includegraphics[width=.6\linewidth]{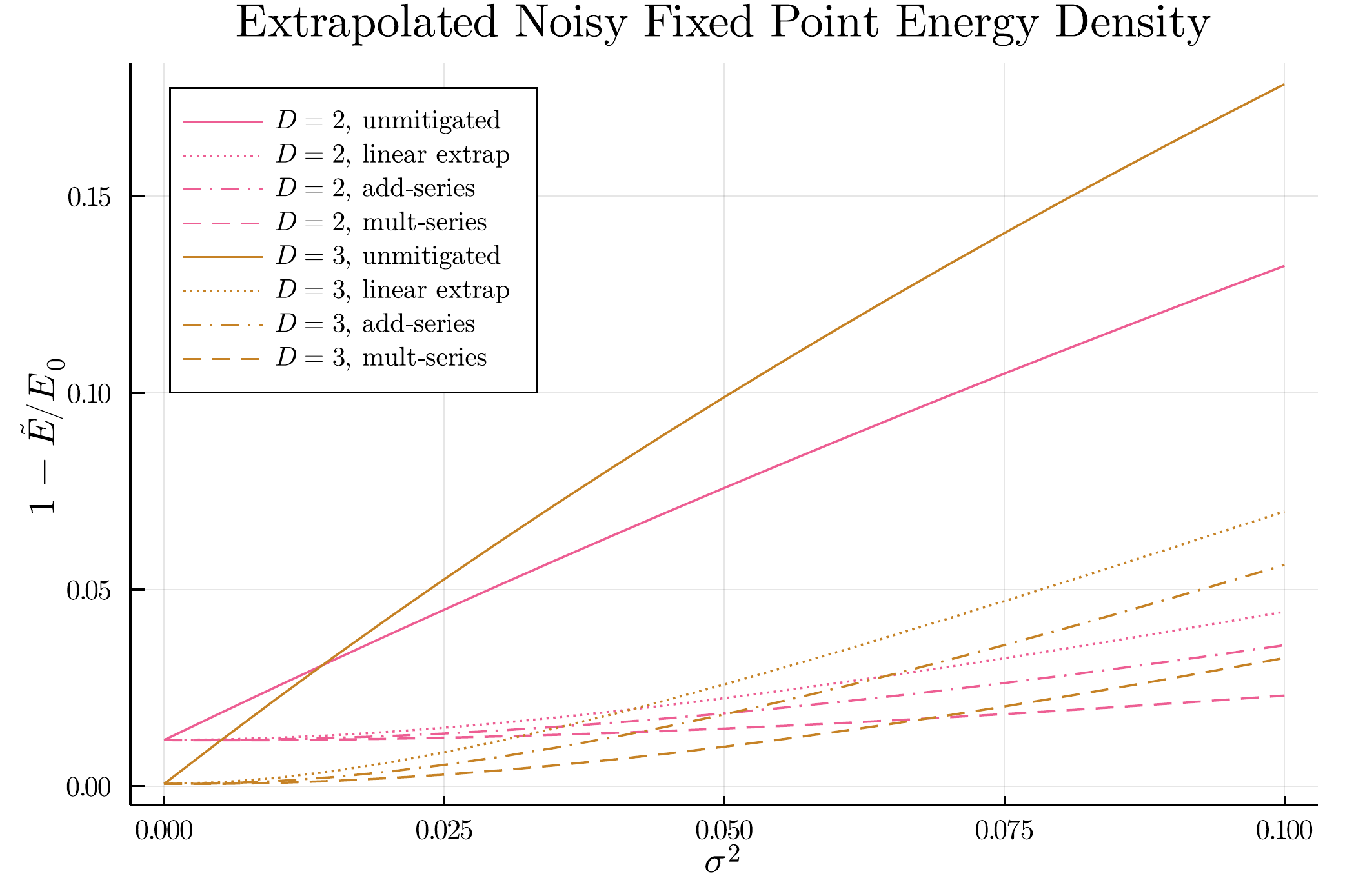}
\caption{Percent error of fixed point energy densities for $D=2$ and $D=3$ channels using the angle imprecision error model and the $\mathcal{C}_1$ gate decomposition. Unmitigated fixed point energies (solid) are considerably improved through zero-noise extrapolation of MS gates using a geometric series error ansatz with additive(dot-dash) or multiplicative(dash) error, both of which outperform standard linear extrapolation of gates in the entire circuit (dotted) that requires much larger gate overhead.}
\label{fig:extrap}
\end{figure}

These model parameters can be fit by amplifying errors in the last and second to last layers of the circuit independently. While this error mitigation technique requires measuring three different versions of the circuit, the gate count overhead for the amplified circuits is not large compared to amplifying the entire circuit. Numerically the scheme also performs much better than full linear extrapolation  applied by amplifying individual gates in the entire circuit (Fig. \ref{fig:extrap}). 
The values of observables obtained through zero-noise extrapolation is not guaranteed to lie between the noisy measurements and the true noiseless value, so the extrapolated energies could lie below the noiseless fixed point values and may not be a variational upper bound to the ground state energy. In the examples considered in our numerics, however, no extrapolation schemes resulted in energies below the noiseless fixed point values.

\section{Experimental Data}

Comparison to other experiments:
Previous experiments using NISQ era ion-traps have prepared critical ground states and thermofield double states for the quantum transverse-field Ising model for small systems using the QAOA variational ansatz \cite{zhu2020generation}. It is known that the critical ground state on $L$ qubits can be prepared exactly with a depth $L/2$ QAOA circuit \cite{Ho_2019}. Since the critical model has polynomially decaying correlations, it is expected that a geometrically local QAOA ansatz would need depth scaling linearly with the system size in order to prepare a good approximation to the ground state. However, with non-local gates it could be possible to lower the required depth to prepare a high fidelity ground state approximation. Using the DMERA ansatz it is possible to prepare polynomially scaling correlations with only a logarithmic-depth circuit by utilizing non-local gates. While the size of our experiment is similarly limited, the DMERA state preparation is better thought of as a finite subystem of the thermodynamic ground state, while the QOAO protocol prepares the exact ground state of a small sized system.

We run our experiment on the Honeywell $\textrm{H\O}$ ion-trap quantum computer. This device has six qubits with global connectivity achieved by bussing the pair of ions to an interaction site when applying the two-qubit gate. Each application of the $n$-qubit channel requires twice as many to implement, so we are limited to three qubit channels but we can apply the channel arbitrary number of times with mid-circuit qubit reset.

The native MS gates of the Honeywell ion-trap are also given in terms of $ZZ$ rotations, rather than $XX$ rotations as used earlier in the paper. Rather than sandwich each MS gate with explicit Hadamard gates to transform them to the other convention, we will implicitly transform the circuits and states to the convention native to the Honeywell trap by interchanging $X$ and $Z$ operators in the experimental implementation. This means our new Hamiltonian will be $H_{I^*} = \sum_i -Z_i Z_{i+1} + Z_{i-1}X_i Z_{i+1}$ and new ancilla will be prepared in the state $\ket{+}$ via a Hadamard transformation on the reset qubits. 

To see the effectiveness of the state preparation algorithm and stability against noise, we measure the expectation value of local observables as a function of the number of iterations of our three qubit channel. With limited resources, we prioritized sampling as many different number of channel iterations as possible to see convergence and stability, rather than repeating individual measurements many times to get better measurement precision. The longest circuits measured had twelve layers of the DMERA channel, each having 120 MS gates, 156 single qubit gates, and 33 qubits recycled mid-circuit, and the average number of samples per measurement was 200.

For initial states we choose products of single qubit states $\ket{\psi_1}$ and $\ket{\psi_2}$ since they can be prepared with high fidelity. The product of the first state $\ket{\psi_1}$ is one of the two mean-field ground states, i.e. a pure product state with the lowest energy, which will actually be below the energy of our noisy fixed point. The second state $\ket{\psi_2}$ is orthogonal to $\ket{\psi_1}$ and serves as a higher energy initial state which will decrease in energy under applications of the noisy channel. These initial states also have opposite signs for the expectations of local $X$ and $Z$ observables which we see converging towards their fixed point values experimentally.

\begin{align}
 \begin{split}
\ket{\psi_1} &= \sqrt{ \frac{3 + 2\sqrt{2}}{6}} \ket{0} - \sqrt{ \frac{3 - 2\sqrt{2}}{6}} \ket{1} \hspace{5mm} E_1 /L = -\frac{32}{27}\\
 \ket{\psi_2} &= \sqrt{ \frac{3 - 2\sqrt{2}}{6}} \ket{0} + \sqrt{ \frac{3 + 2\sqrt{2}}{6}} \ket{1} \hspace{5mm} E_2/L= -\frac{16}{27}
 \end{split}
 \end{align}

\begin{figure}[htbp]
\centering
 \includegraphics[width=.8\linewidth]{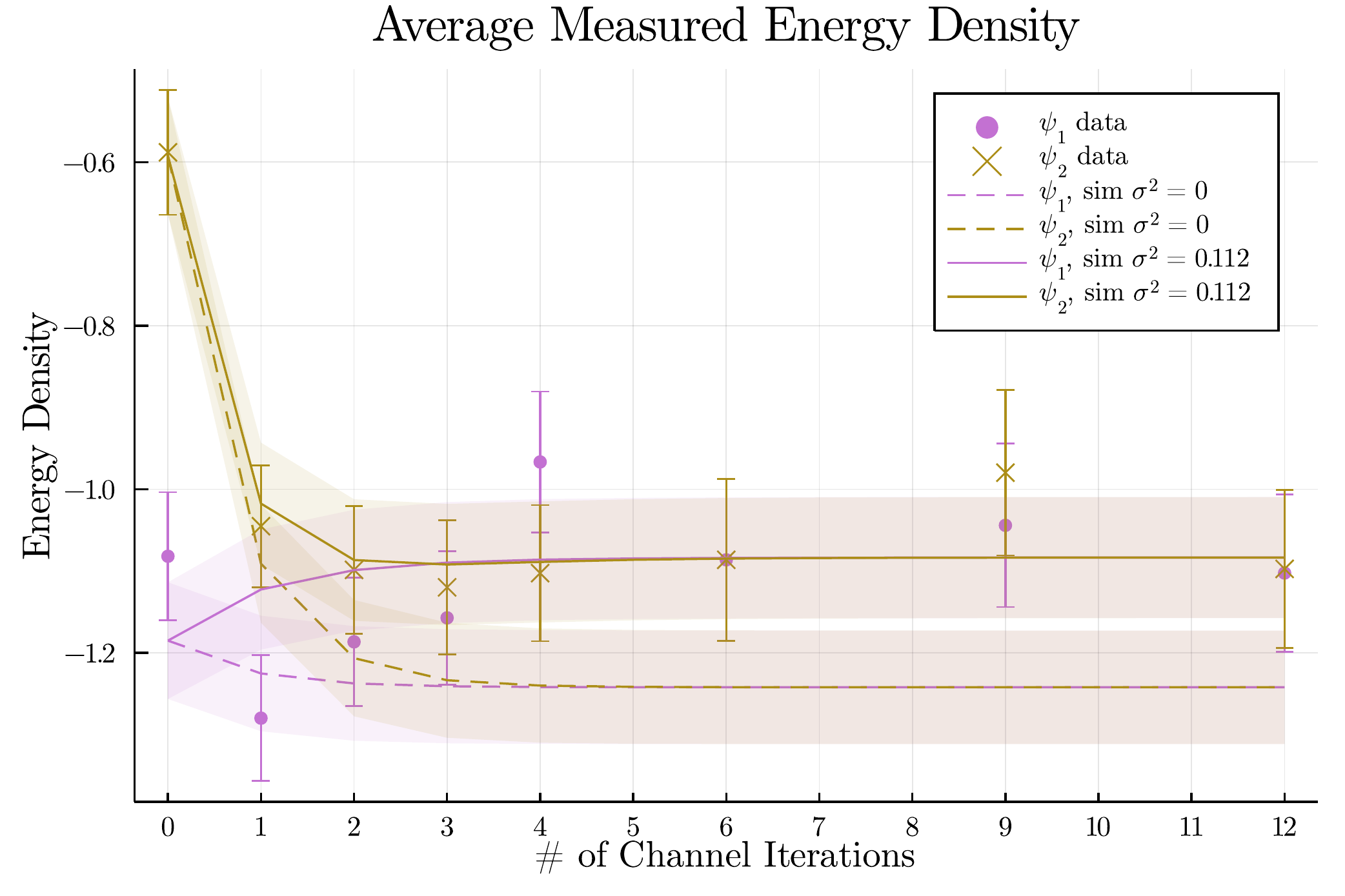}
\caption{Experimental convergence of energy density from initial product states $\ket{\psi_1}$ and $\ket{\psi_2}$. Data from the Honeywell $\textrm{H\O}$ ion-trap computer plotted against simulation curves of noiseless(dashed) and best fit noisy(solid) circuits.}
\label{fig:Edat}
\end{figure}

\begin{figure}[htbp]
\centering
\begin{subfigure}[b]{.48 \textwidth}
 \includegraphics[width=.8\linewidth]{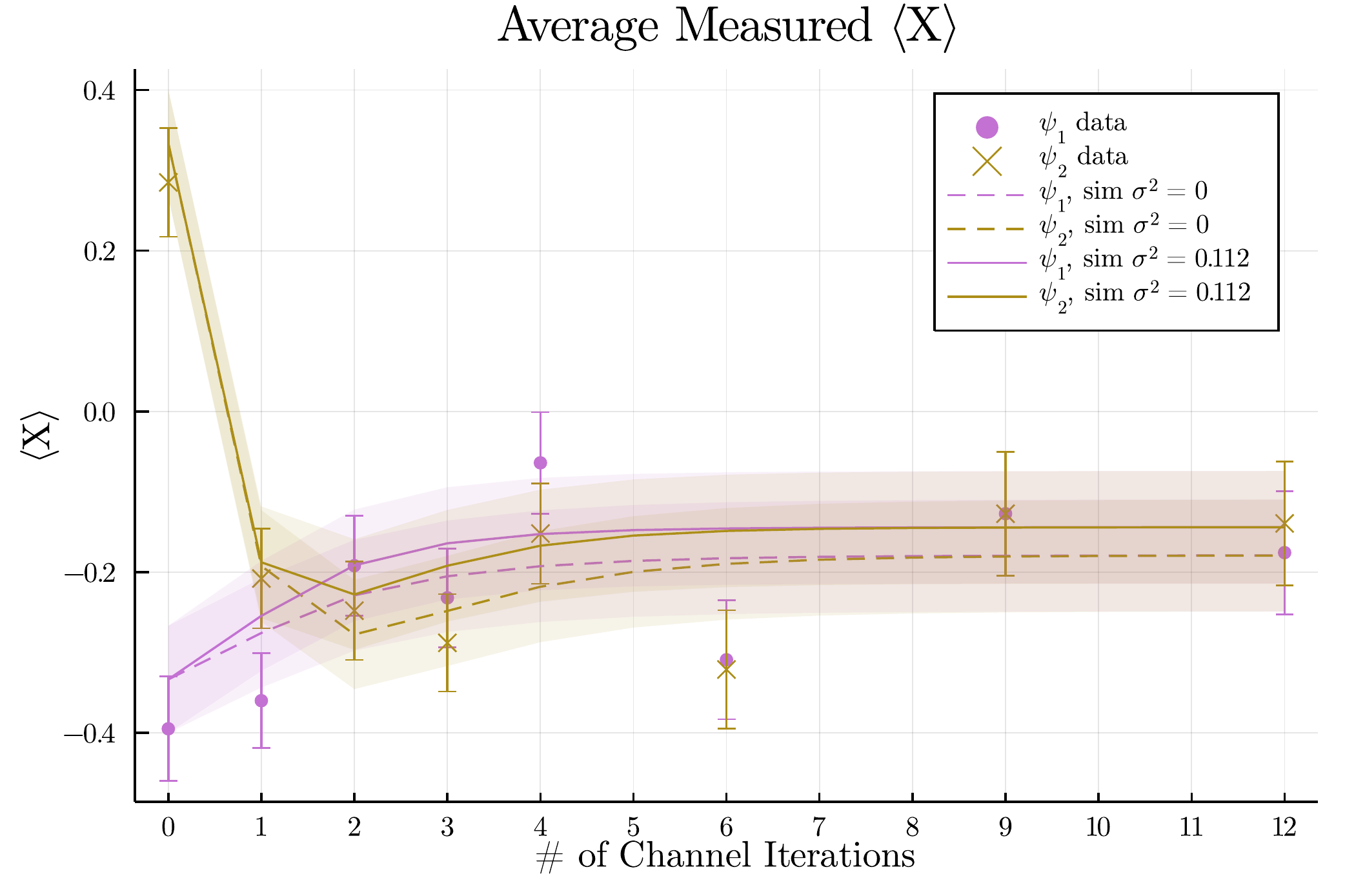}
 \caption{}
 \label{fig:Xdat}
\end{subfigure}
\begin{subfigure}[b]{.48\textwidth}
 \includegraphics[width=.8\linewidth]{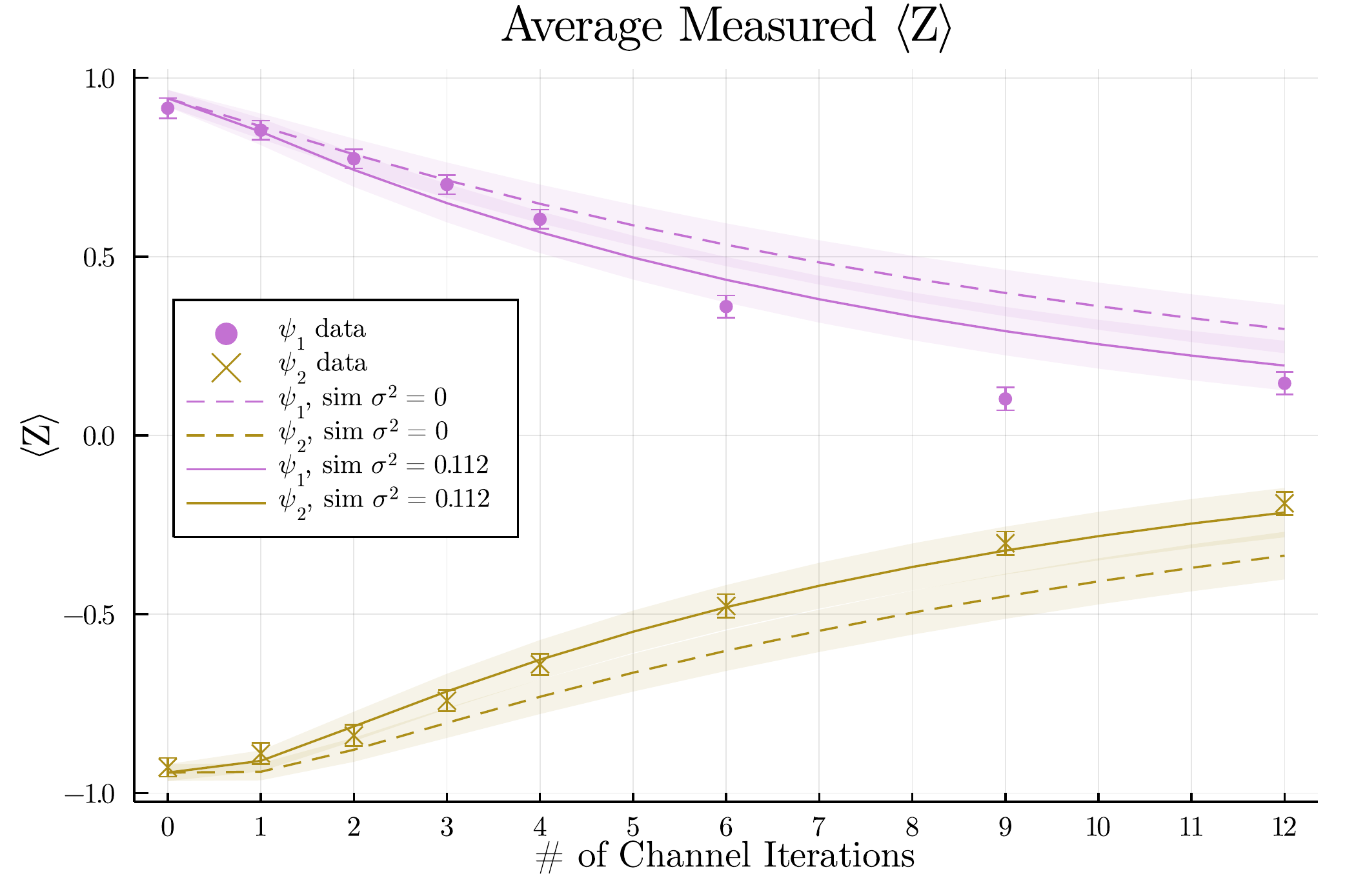}
 \caption{}
 \label{fig:Zdat}
\end{subfigure}
\\
\begin{subfigure}[b]{.48 \textwidth}
 \includegraphics[width=.8\linewidth]{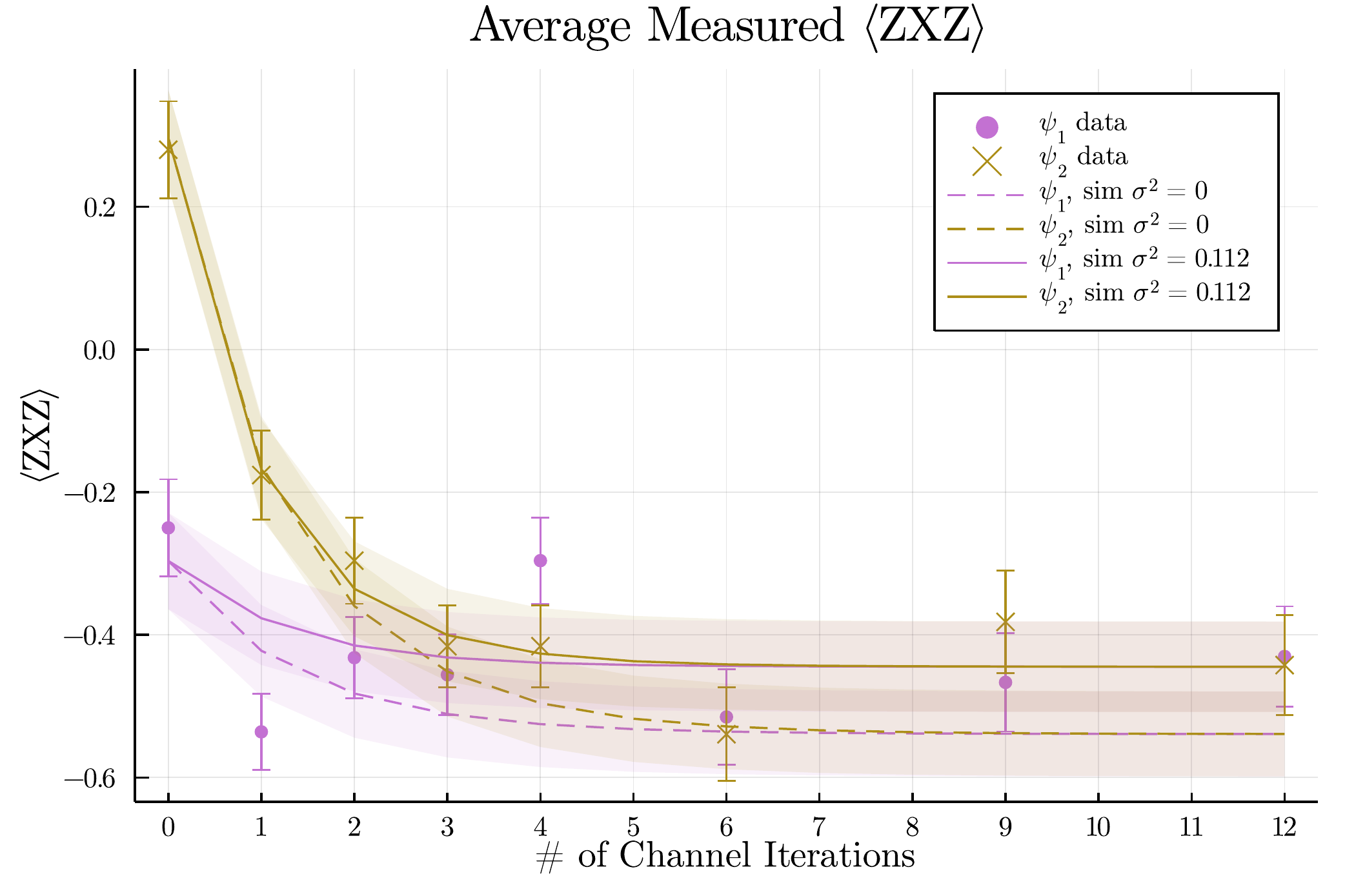}
 \caption{}
 \label{fig:ZXZdat}
 
\end{subfigure}
\begin{subfigure}[b]{.48\textwidth}
 \includegraphics[width=.8\linewidth]{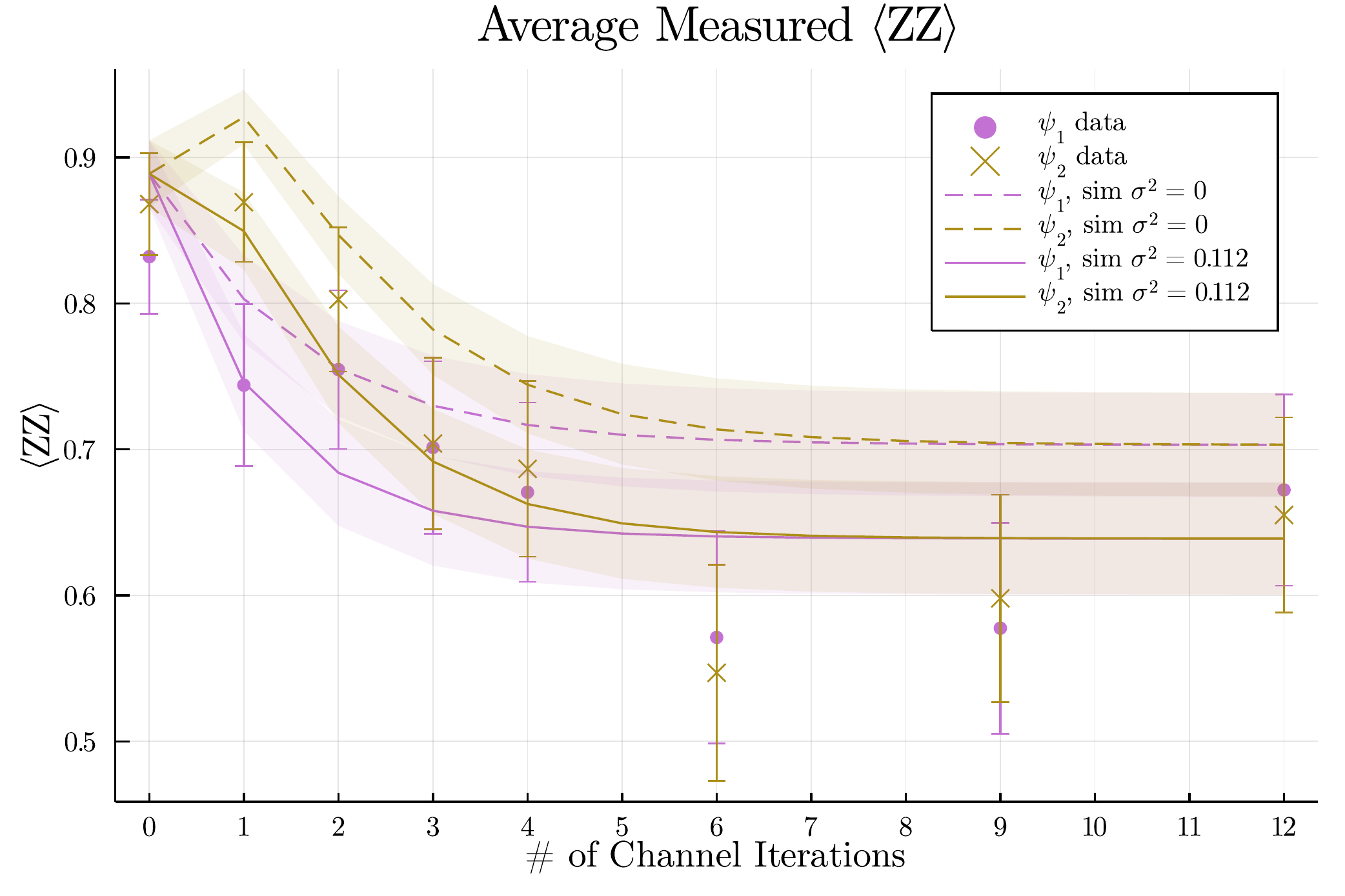}
 \caption{}
 \label{fig:ZZdat}
\end{subfigure}
\caption{Convergence of local expectation values, for single qubit X and Z measurements (a),(b) and multi-spin Hamiltonian terms ZXZ and ZZ (c),(d). Experimental datapoints from the Honeywell $\textrm{H\O}$ ion-trap computer plotted against noiseless and best fit noisy simulation curves, for initial product states $\ket{\psi_1}$ and $\ket{\psi_2}$.}
\label{fig:localDat}
\end{figure}

While there are some outlying data points we find general agreement with the convergence of local observables to somewhat stable values after multiple iterations of our quantum channel. We also see that the convergence rate can vary greatly depending on the exact observable, with the $Z$ expectation converging much slower than either $X$ or Hamiltonian terms $ZZ$ and $ZXZ$. We include in the experimental data for the expectation values of energy density (Fig. \ref{fig:Edat}) and local observables (Fig.\ref{fig:localDat}) the simulated noiseless channel (dashed) and the best fit value of the noise parameter $\sigma^2$ for the angle imprecision model, which we found to be $\sigma^2 = 0.112$. This noise model corresponds to a fixed point energy density of $-1.08$, 15.2\% higher than the true ground state energy density of $-4/\pi$.

\section{Conclusions}
We were able to numerically and experimentally demonstrate the convergence and noise-resilience of local dissipative circuits that prepare subregions of renormalization fixed point states, specifically using circuits derived from the unitary wavelet transforms in \cite{ERwav} for the ground state of the critical Ising model. Numerical noise models of gate imprecision and local depolarizing noise support a linear scaling of the fixed point observables in the local error parameters $\sigma^2$ and $p$, and the channel depth $D$. We also showed evidence that different choices of gate decomposition may greatly affect the noise susceptibility for some kinds of noise.

Error mitigation techniques such as zero-noise extrapolation are also seen to dramatically improve the mitigated error rate of observables with relatively little gate overhead when using extrapolation schemes designed using knowledge about the error propagation for a dissipative channel. These effectively lower noise levels achieved from classical post-processing can be leveraged to obtain the approximation accuracy of larger-depth circuits even in regimes where their raw output is noisier than shallower-depth circuits.

We are interested in how similar DMERA circuits which prepare states in a noise-resilient way could be adapted for models away from a critical point, for thermal states, or for non-integrable models. It has also yet to be seen how the scaling with channel depth and benefits of our error mitigation scheme hold up for real experimental data, but which may all be tested relatively soon on NISQ devices. 

There may also be interesting connections with our work and holographic models of quantum gravity and error correction. This relation could arise from the interpretation of MERA-like circuits as a description of the AdS/CFT correspondence\cite{Swingle_2012} and the recognition of error correcting properties in the structure of the duality \cite{Pastawski_2015, Almheiri_2015, Kim_2017}. Alternatively, the robustness of our circuit may be interpreted holographically in terms of the dilution of noise in an expanding universe \cite{DMERA, Bao_2017}.

\section{Acknowledgements}

We gratefully acknowledge support
from the Department of Energy under award number DE-SC0019139. We also thank Ning Bao, Aniruddha Bapat, and Ruben Andrist for helpful discussions and comments.

\bibliographystyle{unsrt} 
\bibliography{refs.bib}

\end{document}